\documentclass{article}
\usepackage{graphicx}
\usepackage{amsmath}
\begin{document}

\title{Resistive threshold logic}
\author{A. P. James, L.V.J. Francis and D. Kumar}
\date{}
\maketitle

\begin{abstract}
    We report a resistance based threshold logic family useful for mimicking brain like large variable logic functions in VLSI. 
A universal Boolean logic cell based on an analog resistive  divider and  threshold logic circuit is presented. The  resistive divider is implemented using memristors and provides output voltage as a summation of weighted product of  input voltages. The output of resistive divider is converted into a binary value by a threshold operation implemented by CMOS inverter and/or Opamp. An universal cell structure is presented to decrease the overall implementation complexity and number of components. When the number of input variables become very high, the proposed cell offers advantages of smaller area and design simplicity in comparison with CMOS based logic circuits.  
\end{abstract}

\section{Introduction}
%
%
%
%
Logic gates implement boolean algebraic expressions obtained from truth tables. Increase in functional requirements of  digital IC's such as in microprocessors and ASIC's results in complex logic state implementations. A complex set of logic states when represented as a truth table would have large number of input and output variables. As the number of input variables increases, it is often not possible to manually reduce the boolean logic expressions to reduce the number of components required for its implementation. The most common approach to reduce the number of components
required with a large number of variables is by using logic minimisation
based on prime implicant logics. Technique such as Karnaugh map\cite{ref_1}, QuineMcCluskey\cite{ref_3},  Petrick's method, Buchberger's algorithm \cite{ref_7} and Espresso minimization algorithm \cite{ref_8}, are the widely used approaches.  However, when the number of inputs increases significantly, logic minimisation methods become inefficient. In addition,  implementations using existing logic families become challenging as they are often restricted by the gate delays, the number of inputs and the number of components. 

The common approach employed to implement boolean algebra with a large number (\textgreater10)\ of  variables, is to apply the minimization techniques for standard gates with a limited number of inputs (\textless10). This always results in more number of circuit components than that was possible with gates that could support as many number of inputs as the number of variables. In addition to this issue, the number of components required to implement a  gate vary from one boolean logic to another, which results in increased structural complexity and results in increased investment in production scale verification and testing cycles.   

Generic digital circuits such as a single  2$^n$ to 1 multiplexer can be used to implement $n$-input boolean logic function in canonical sum-of-products form. As the number of inputs to the multiplexer increases, a typical AND-OR logic would have large number of inputs per gate for its implementation. In order to implement large variable boolean logic functions such as using multiplexers, we introduce the concept of resistance threshold logic that minimises number of components and design complexity. The proposed resistive threshold logic is made up of a resistive divider  and a threshold logic circuit.  The idea of such an analog-binary cell is inspired from the implementation challenges of the long established theory and practices of neuron cell modelling and logic circuits \cite{ref22}. Conventional neuron inspired logic gate implementations\cite{20} are complex due to the requirements of multi-valued weights and neuron like threshold functions. In addition, they  fail to meet the original aim of having large input logic gates useful for mimicking brain like logic functions. In contrast, the resistive threshold logic is aimed to be simple in structure having the ability  to realise large variable   logic functions,  and is intended to be used as a new standard cell universal logic family with a possible ability to mimic brain logic.  

%

\section{Proposed Cell}
\begin{figure}[ht]
\centering
\includegraphics[width=50mm]{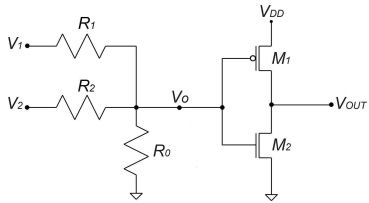}
\caption{The circuit diagram of the proposed resistive divider boolean logic cell that consists of a two input resistive divider and a variable threshold CMOS inverter is presented.}
\label{fig1}
\end{figure}

The proposed logic cell shown in Fig. \ref{fig1} consists of a resistive divider and a variable threshold inverter.  In contrast to the earlier reported work on cognitive memory network \cite{ref_10}, in this work, we propose a  significantly different configuration, implementation  and application of the structurally similar and conceptually different cell.  The input  to the resistive divider are the digital values that can be equated to the logic inputs of a digital logic gate. Based on the output of the resistive  divider and  a predefined inverter threshold, we propose to implement the basic boolean  logic functions. The selection of the threshold and the use of resistive logic in designing a generalized logic cell is the primary contribution of this research.


An $N$-input resistance divider circuit consists of $N$ input resistors $R_i$ and one reference resistor $R_0$. The output voltage $V_0$ for $N$-input voltages $V_i$ can be represnted as 
$V_0={\sum_{i=1}^N \frac{V_i}{R_i} }/{(\frac{1}{R_0} +\sum_{i=1}^N \frac{1}{R_i} )} $.
The inputs  $V_i$ have either of the two logical levels  $V_H$ or  $V_L$, representing a binary logic [1,0]. We keep equal values to $R_{i}'s$  and $R_{0}=mR_{i}$, which results in:
$V_0=\frac{\sum_{i=1}^N {V_i}}{\frac{1}{m}+{N}}$.
                             
A straight forward approach to implement  resistors is by using semiconductor resistors. Semiconductor resistors consist of a resistive body that is surrounded by an insulator often developed over a substrate, and two terminal contacts implemented using conductive metallic strips. The value of semiconductor resistance can be obtained from the expression, $\frac{\rho L}{x_jW}$, where  $\rho$ is the resistivity,  $L$ is the length,   $x_j$ is the layer thickness and $W$ is the width of the resistive body.   

\begin{figure}[ht!]
\centering
\includegraphics[width=50mm]{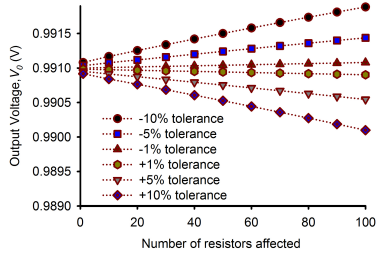}
\caption{The impact of change in input resistance on the output voltage $V_0$ of the resistive divider is 
graphically illustrated. The results are demonstrated for 100 input resistive divider, with each line showing the relative change in $V_0$ for the corresponding number of resistors are uniformly perturbated within a $\pm 10\%$ tolerance level of resistor values. Note: here we keep $V_i=1$.}
\label{fig2}
\end{figure}

A concern while using  resistance devices (such as semiconductor resistors) is the  impact of   change in resistance value due to  second order implementation effects, such as improper junctions and defects. Figure 2 shows a simulated study of the impact of change in resistance values on the output voltage of a  resistive divider circuit. It is assumed here that the changes in the  resistor values are limited within a tolerance level of $\pm 10\%$ of the actual resistive values.
It can be seen that a maximum of $\pm10\%$ resistive values introduces only about $.0894\%$ change in output voltage, which makes the practical implementation of the resistive divider feasible even under realistic conditions. While using semiconductor resistors, when the number of inputs increase, the leakage current through the semiconductor resistance becomes prohibitively high. This drawback is overcome by replacing semiconductor resistors with memristors \cite{ref_14}, which has negligible amount of leakage current.
 
The proposed resistive divider circuit uses the memristor modeled by HP \cite{ref_14}. The device has a thin film of titanium dioxide (TiO$_2$) sandwiched between two platinum terminals. The titanium dioxide layer is doped on one side with oxygen vacancies, TiO$_{2-x}$. The doped region has lower resistance than that of the insulated undoped region. The boundary between doped and undoped region determines the effective resistance of the device. Let $D$ be the total width of the TiO$_2$ layer and $W$ be the width of the doped TiO$_2$ layer. When a positive voltage is applied at the doped side, the oxygen vacancies moves towards the undoped region, increasing the width of the doped region, $W$ and hence the effective resistance of the memristor decreases. The effective resistance $M_{eff}$ of the memristor is 
$M_{eff}=\frac{W}{D}R_{ON}+(1-\frac{W}{D})R_{OFF}$,
where, $R_{ON}$ (=1 k$\Omega$) is the resistance of the memristor if it is completely doped and $R_{OFF}$ (=100 k$\Omega$) is the resistance of the memristor if it is undoped. When input voltage is withdrawn or when there is no potential difference between the terminals, the memristor maintains the boundary between the doped and undoped region, since the oxygen ions remain immobile after removal of the input voltage. Thus the resistance will be maintained at the same value before withdrawing the input voltage. From the equation, $i=\frac{v}{M(q)}$ \cite{ref_15}, where $v$ and $i$ are the voltage and current across the memristor, and $M(q)$ is charge dependent resistance of the memristor, we can see that when the voltage difference across the memristor is $0$, the current through the memristor is $0$. If there is a reverse potential across the memristor, the width of the undoped region increases, resulting in an increase in the effective resistance of the memristor. This high resistance will block the reverse leakage current through the memristor. When the number of inputs increases, the collective forward current through the circuit does not increase significantly, since the effective resistance in the memristor is constant. Table \ref{table1} shows the effect of increase in number of inputs on the collective current flowing through the circuit.

\begin{table}[ht]
\centering
\caption{Effect of increase in number of inputs on the forward current flowing through the memristor in the circuit.}
\begin{scriptsize}\begin{tabular}{p{2cm}p{2cm}p{2cm}}\hline\hline
Number of inputs & Current through a single memristor & Current through the potential divider circuits \\\hline
2&3.33$\mu$A&6.66$\mu$A  \\ 
10 & 0.909$\mu$A & 9.09$\mu$A \\
100 & 0.99099nA & 9.90099$\mu$A \\
\hline\hline
\end{tabular}
\end{scriptsize}\label{table1}
\end{table}

\begin{table}[ht]
\centering
\caption{Truth Table of Two Input Resistive Divider Logic Cell When Used as NAND and NOR Gates }
\begin{scriptsize}
\begin{tabular}{p{1cm}p{1cm}p{1cm}p{1cm}p{1cm}}\hline\hline
\multicolumn{2}{p{2cm}}{Input Voltage ($V_i$)} & Output Voltage &  NAND$^a$& NOR$^b$ \\
$V_1$ & $V_2$ & $V_0$ &  &   \\\hline 
$V_L$ & $V_L$ & $\frac{2V_L}{3}$ &  $V_H$ & $V_H$ \\
$V_L$ & $V_H$ & $\frac{V_L+V_H}{3}$  & $V_H$ & $V_L$ \\
$V_H$ & $V_L$ & $\frac{V_L+V_H}{3}$  & $V_H$ & $V_L$ \\
$V_H$ & $V_H$ & $\frac{2V_H}{3}$  & $V_L$ & $V_L$ \\
\hline\hline
\multicolumn{5}{p{6cm}}{$^a$ NAND threshold range $\frac{V_L+V_H}{3}< V_{th}<\frac{2V_H}{3}$}\\
\multicolumn{5}{p{6cm}}{$^b$ NOR threshold range $\frac{2V_L}{3}<V_{th}<\frac{V_L+V_H}{3}$}\\
\end{tabular}
\end{scriptsize}
\label{table2}
\end{table}

Table \ref{table2} shows the truth table of the two input resistive divider logic cell, that implements the NAND and NOR gates using a predefined inverter threshold $V_{th}$. Assuming that $V_{dd}=1V, V_H=1V, V_L=0V$ it is clear from Table \ref{table2} that if the threshold voltage of the inverter is set between $0V$ and $1/3V$, the cell will work as NOR logic and if it is between $2/3V$ and $1/3V$ the cell will work as NAND logic. That means by varying the threshold voltage of the inverter, NAND and NOR logic can be implemented using a single cell.  
In general, the range of threshold voltage, $V_{th}$ of
 NOR gate is  $\frac{NmV_{L}}{1+Nm} \le  V_{th} \le   \frac{(V_H+(N-1)V_L)m}{Nm+1} $    , and  NAND gate is, $\frac{m(V_L+(N-1)V_H)}{(Nm+1)}\le  V_{th} \le  \frac{mNV_H }{Nm+1}$.
 To find the $m$ value, the lower limit of NAND gate threshold range $(\frac{m(V_L+(N-1)V_H)}{(Nm+1)})$ is equated to $\frac{V_H+V_L}{2}$. Now if we assume $V_L$  as $0V$ then we get the $m$ value as$ \frac{1}{N-2}$ and  we can say that  the threshold voltage of NAND gate must be between $\frac{V_H+V_L}{2}$ and $\frac{mNV_H }{Nm+1}$.


The threshold voltage of the MOSFET is dependent on several parameters such as substrate bias voltage $V_{bs}$,  the surface potential $\phi_s$,  and substrate doping concentration \cite{ref_16}. The threshold voltage $V_{tn}$ of the MOSFET can be varied by changing its substrate bias, $V_{bs}$. The dependence of substrate bias and the threshold voltage 
is expressed as,
$V_{tn}=V_{tn0}+K_1(\sqrt{\phi_s-V_{bs}}-\sqrt{\phi_s})+C$
, where, $V_{tn0}$ is the zero bias threshold voltage,  the surface potential  $\phi_s=2\frac{k_BT}{q}\ln(\frac{N_a}{n_i})$, $K_1 $ is a parameter derived by considering non-uniform doping and short channel effects $K_1=\gamma_2-2K_2\sqrt{\phi_s-V_{bm}}$ where $K_{2}=\frac{(\gamma_1 - \gamma_2)(\sqrt{\phi_s-V_bx}-\sqrt{\phi_s})}{2\sqrt{\phi_s}(\sqrt{\phi_s-V_{bm}}-\sqrt{\phi_s})+V_{bm}}$ $\gamma_1$ and $\gamma_2$ are body bias coefficient when substrate doping concentration are equal to $N_{ch}$ and $N_{sub}$ respectively. $\gamma_1=\frac{\sqrt{2q\epsilon_{Si}N_{ch}}}{C_{ox}} , \gamma_2=\frac{\sqrt{2q\epsilon_{Si}N_{sub}}}{C_{ox}} $ and $V_{bm}$ is the maximum substrate bias voltage. And $C$ shows the effect of narrow channel on threshold voltage. 
 The threshold voltage of the inverter can be represented as, $V_{th}=\left({(V_{tn}+(V_{DD}-|V_{tp} | ))\sqrt{\frac{\mu_{p}W_{p}}{\mu_{n}W_{n}}}}\right)/\left({1+\sqrt{\frac{\mu_{p}W_{p}}{\mu_{n}W_{n}}}}\right)$, which shows the role of the threshold voltages of the MOSFETs in determining the threshold of the inverter.




\begin{figure}[ht!]
\centering
\includegraphics[width=50mm]{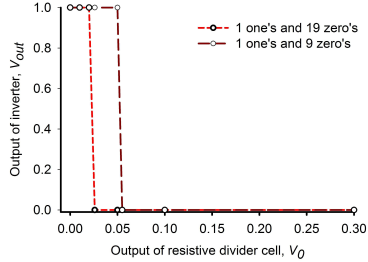}
\caption{The relation between Output voltage of the inverter and Output voltage of the resistive divider, for 10 input and 20 input boolean logic, when it is working as a NOR gate is shown}
\label{fig3}
\end{figure}

\begin{figure}[ht!]
\centering
\includegraphics[width=80mm]{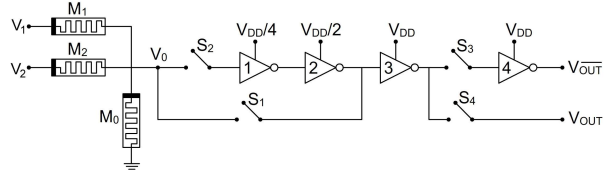}
\caption{The circuit diagram to implement  NAND, NOR, AND, OR and NOT logic functions consisting of memristive resistance divider and CMOS inverters with three different power supply values.}
\label{fig4}
\end{figure}

 Fig. \ref{fig3} shows the relationship between the output voltage of the resistive divider cell (input to the inverter) and the output voltage of an inverter, for 10 input and 20 input situations, when the cell is working in NOR logic. $V_{0}$ value when the inputs are $V_1=1$ and $V_{2}=V_{3}=..V_{10}=0$ is $0.0556V$, and when $V_{1}=V_{2}=..V_{10}=0$ is $0$, so the threshold voltage of the inverter must be between $0$ and $0.0556$, to work as a NOR logic. Similarly for 20 input boolean logic, the threshold voltage of the inverter must be between $0 $ and $0.026$. This shows that if the threshold voltage of the inverter can be lowered to a very small value we can implement resistive threshold logic with large number of inputs.\newline
In order to reduce the threshold voltage, here we introduced three inverters with three different $V_{DD}$'s. Fig. \ref{fig4}  shows  a universal gate structure which can be used to implement AND, NAND, OR, NOR and NOT logic. For the cell to work as a NAND logic, the switches $S_1$ and $S_4$ are closed, and the output is taken from $V_{out}$. So in this case, three inverters will be enabled. To implement AND logic, the switches $S_1$ and $S_3$ are closed, and the output is taken from $\overline{V_{out}}$. For the AND logic, two inverters need to be enabled. If the switches $S_2$ and $S_4$ are closed, we get a NOR logic from $V_{out}$, here only one inverter has to be enabled. If both $S_2$ and $S_3$ are closed, OR logic can be implemented, here two inverters are used. The approach shown in Fig. \ref{fig4}, demonstrates the concept of generalization of resistive threshold logic cell to implement the most basic boolean logic functions. To maintain practical relevance of the approach all the results reported are based on device parameters from  0.25$\mu m$  TSMC process. Note that as  $V_{DD}$ decreases  $V_{th}$ also decreases. When $V_{DD}$ changes the $V_{GS}$  of PMOS in the CMOS inverter will also change. As a result, in the case of the proposed cell with 10 inputs, the PMOS will be in cut off state when the input condition is  $V_1=1$ and $V_{2}=V_{3}=..V_{10}=0$ and we get a low level output from the 1st inverter.  Since the 1st inverter can only provide a high value of $0.25V$, we use other two inverters in order to get a high value of $1V$.
The working of the proposed cell in Fig. \ref{fig4} as a NAND or NOR gate purely rests on the values of $V_{tn}$ and $V_{th}$ of the inverter, for a given number of inputs.

\begin{figure}[ht!]
\centering
\includegraphics[width=50mm]{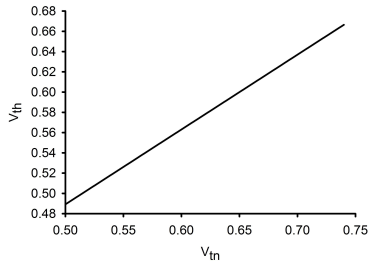}
\caption{A graph indicating the dependence of threshold voltage of the CMOS inverter and threshold voltage of the NMOS. The threshold values shown in the graph is a result of changing the number of inputs from 3 to 100 and calculating the minimum inverter threshold voltages required to implement the circuit as a NAND gate}

\label{fig5}
\end{figure}

If  $V_H$ is set as $1V$ and $V_L$ as $0$, then the threshold voltage $V_{th}$ range for NAND gate must be between $0.5V$ and the $V_0$ value obtained  when all inputs are $V_H$.  Figure \ref{fig5} shows the relationship that exists between $V_{tn}$ and $V_{th}$ to implement the proposed cell as NAND gate, as the number of inputs changes from 3 to 100.  For each  number of inputs the  $V_{th}$ is calculated  for a particular $V_{tn}$ and with a fixed $V_{tp}$, $W_p$, $\mu_p$, $W_n$, $\mu_n$ and $V_{DD}$ values. For a given number of inputs the threshold voltage is above $0.5V$, so by using a single inverter with $V_{DD}$ as $1V$, NAND logic can be implemented. That means NAND logic can be implemented using the proposed cell with one inverter such as in Fig. \ref{fig1}. Using three inverters with different $V_{DD}$, a 100 input NOR logic can be realised. For implementing NOR logic, for larger number of inputs, the threshold voltage of the inverter circuit has to be reduced to a very low value. This  problem can be overcome by boosting the signal, using an Opamp amplifier, before applying to the inverter. Table \ref{tabl3} shows the  leakage power and the  spectral noise due to Johnson, shot and flicker noise in  multi-$V_{DD}$ logic proposed in Fig 4. The the maximum noise levels are very low (ie in nV) relative to signal reference of 1V range.  

\begin{table}

\centering

\caption{Leakage power and noise spectral density for 100 input gate proposed multi-$V_{DD}$ gate configuration in Fig 4}
\begin{scriptsize}
\begin{tabular}{|p{4cm}|c|c|c|c|}\hline
 Performance measure& NAND & AND & NOR & OR \\\hline
 Noise spectral density per unit square root bandwidth ($nV/Hz^{1/2}$)& 7.94 & 9.75 & 75.71 & 10.15 \\\hline
Leakage power ($nW$) & 0.014 & 0.017 & 0.967 & 0.971\\\hline
\end{tabular}\end{scriptsize}
\label{tabl3}
\end{table}

\begin{figure}[ht!]
\centering
\includegraphics[width=70mm]{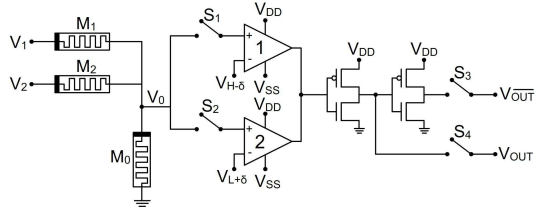}
\caption{The universal gate structure that implements NAND, NOR, AND, OR and NOT logic functions using memristive resistance divider and Opamp threshold circuit.}

\label{fig6}
\end{figure}

The universal circuit in Fig \ref{fig4} is modified to incorporate Opamp threshold logic as shown in Fig. \ref{fig6}.  The threshold logic when implemented using  Opamp \cite{ref_17}, offers the advantage of scalability over increase in number of inputs. The Opamp is designed using 8 MOSFETs and in the same technology as that of the CMOS NOT gate. The Opamp reference voltage for NOR logic, $V_{REF}$ is fixed as $V_{L}+\delta$ and for NAND logic, $V_{REF}$ is fixed as $V_{H}-\delta$, where $\delta$ is small voltage defined to ensure the bounds of $V_{th}$. The Opamp  shifts the voltage to a high value or low value depending on the input voltage, $V_{0}$. It also acts as a buffer helping to isolate the inputs from the output enabling realistic implementations of very large of inputs per gate.

\subsection{Comparisons}

\begin{figure}[ht!]
\centering
\includegraphics[width=50mm]{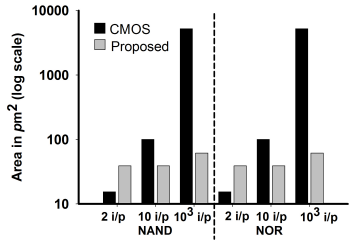}
\caption{The bar graph shows the area comparison of  CMOS with that of  Resistive Threshold Logic (with Opamp threshold circuit, Fig. 6), using NAND and NOR gate implementations.}
\label{fig7}
\end{figure}

 Fig. \ref{fig7} indicates the area required to implement NOR and NAND universal logic gates for 2, 10, and 1000 input logic gates implemented using CMOS logic, and that using the resistive threshold logic. In implementing CMOS logic the maximum number of inputs per gate is taken as 5. The Fan in of the proposed cell using Opamp  is very high ($=14.498\times 10^6$), indicating that we can implement a large variable boolean logic using a single resistive divider cell. For increasing number of inputs, the proposed cells contain lesser number of components and area, when compared to the CMOS logic. Since CMOS based logic gates  are practically limited to small number of inputs, we have used a layered combination of 5 input gates to implement gates with 10 or more inputs. Table \ref{table3} compares the power dissipation of the proposed logic with that of CMOS logic for NAND  and NOR gates. CMOS gates dissipates lesser power as against its memristive counterparts.  The use of low power memristive devices\cite{ref_18} would be required to reduce the power dissipation.   Table \ref{table4} shows the comparison of the noise margin of the logic families for single input NAND and NOR logic, indicating that the proposed logic has comparable noise tolerance levels to that with the existing techniques.  In addition, the averaging nature of the potential divider can further help to increase the noise tolerance levels than specified through noise margins. Table \ref{table5} shows a comparison of propagation delay when a square pulse with 40$\mu$s time period and 50\% duty cycle is applied.  The resistive threshold logic shows better response when the number of inputs become very high, and when with lower number of inputs show comparable delays. 

\begin{table}[ht!]
\centering
\caption{ Comparison of the Resistive Logic with CMOS Logic}
     \begin{scriptsize}    
                
                \begin{tabular}{p{1.9cm}p{1.cm}p{1.cm}p{1.cm}}\hline \hline
     Logic family$^{a}$ & Logic function &\multicolumn{2}{c}{Power Dissipation}\\\cline{3-4}
&&10 i/p&100i/p\\
        \hline 
CMOS logic &&0.009nW&0.036nW \\
Resistive logic (Opamp threshold)& NOR & 10.6$\mu$W&11.49$\mu$W\\
                \hline
CMOS logic &&0.062nW&0.753nW\\
Resistive logic (Opamp threshold)&NAND&9.2$\mu$W&10.09$\mu$W\\
                \hline

\multicolumn{4}{p{6cm}}{$^{a}$The technology size of all the components in the circuit is kept same for all the gates for fairness in comparison. }
  \end{tabular}
        
        \end{scriptsize}

  \label{table3}
\end{table}
 
 As the resistance elements does not significantly introduce the delay with increase in number of inputs, a large number of inputs (\textgreater 100)  is practically possible for the proposed cell. In contrast with the existing technologies that are practically limited to about 5-10 inputs per gate, the ability of the proposed resistive threshold logic to handle large number of inputs  reduces the complexity of the design and layout of the large variable digital circuits.

\begin{table}
\centering
\caption{Noise margin of different logic families}
\begin{scriptsize}\begin{tabular}{p{2.5cm}p{1cm}p{1cm}p{1cm}p{1cm}}\hline\hline
Logic families & \multicolumn{2}{c}{NAND}& \multicolumn{2}{c}{NOR} \\\cline{2-5}
& NM$_L$ & NM$_H$  & NM$_L$ & NM$_H$\\\hline
CMOS&0.363V& 0.587V& 0.233V&0.616V\\
Pseudo NMOS & 0.429V & 0.413V&0.276V&0.461V\\
Domino CMOS & 0.407V &0.376V &0.104V&0.43V\\
Resistive logic  & 0.369V & 0.558V&0.132V&0.777V\\
\hline\hline
\end{tabular}
\end{scriptsize}\label{table4}
\end{table}

\begin{table}
\centering
\caption{Propagation delay of different logic families for different number of inputs}
\begin{scriptsize}\begin{tabular}{p{2cm}p{.5cm}p{.5cm}p{.5cm}p{.5cm}p{.5cm}p{.5cm}}\hline\hline
Logic families & \multicolumn{3}{c}{NAND delay}&\multicolumn{3}{c}{NOR delay}\\\cline{2-7}
& 3i/p & 10i/p&1000i/p&3i/p & 10i/p&1000i/p \\\hline
CMOS&0.47$\mu$s& 0.54$\mu$s& 0.65$\mu$s&0.50$\mu$s&0.52$\mu$s&0.66$\mu$s\\
Pseudo NMOS & 0.48$\mu$s & 0.60$\mu$s&0.85$\mu$s&0.51$\mu$s&0.58$\mu$s&0.72$\mu$s\\
Domino CMOS & 0.48$\mu$s &0.51$\mu$s&0.75$\mu$s&0.51$\mu$s&0.58$\mu$s&0.75$\mu$s\\
Resistive logic (Opamp threshold) &0.45$\mu$s&0.45$\mu$s&0.45$\mu$s&0.60$\mu$s&0.60$\mu$s&0.60$\mu$s\\
\hline\hline
\end{tabular}\end{scriptsize}
\label{table5}
\end{table}

\subsection{Example Circuits}

The proposed logic  is compared with the  CMOS implementation using a 16 bit adder and a 16x1 MUX. The simulation were performed in spice using feature size of 0.25$\mu$m TSMC process BSIM models and HP memristor model.  A ripple carry adder without applying  reduction technique is implemented using 16 single bit adders. The single bit adder
require 3 NOT, 3 two input AND, 1 three input OR, 4 three input AND and 1 four input
OR gates. Hence, a  total of  48 NOT, 24 AND, 16 OR, 64 AND and 16 OR gates are required for the 16 bit adder.  Figure 8 shows an example of   16th output bit of the adder simulated using input pulses with initial start delay of 10$\mu$s, rise and fall time of 5$n$s, and ON period of either  20$\mu$s or 10$\mu$s  with 50\% duty cycle. 
\begin{figure}[ht]
\centering
\includegraphics[width=80mm]{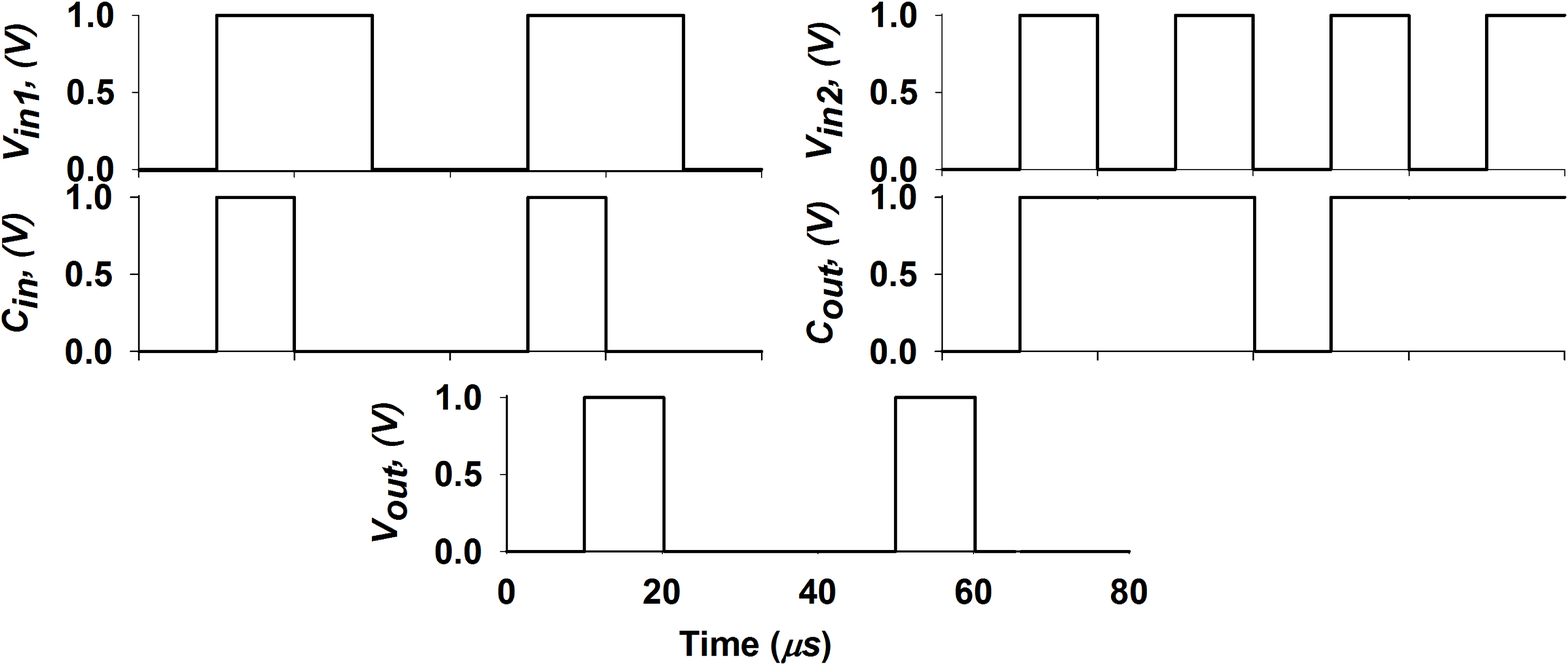}
\caption{The signal output of the 16th bit of the designed ripple adder using the proposed resistive threshold logic. $V_{in}$'s is the inputs,$C_{in}$ and $C_{out}$ is the input and output carry, and $V_{out}$ the output sum bit. }
\end{figure}

The 16 bit MUX when using the proposed logic required 16 input OR gate and 5 input AND gates, while CMOS logic required  2, 4 and 5 input AND/OR gates. In the case of adder, CMOS logic has lesser area in comparison to the resistive threshold logic, while in 16x1 MUX implementation proposed logic result in lesser area when compared to CMOS logic. Table \ref{table71} demonstrates  that when the number inputs for the AND and OR gates are increased, the proposed logic require lesser area than its CMOS counterpart. Power dissipation on the other hand is higher for the proposed logic due to higher forward currents in memristor  as compared with CMOS. This issue can be addressed by using low power memristors \cite{ref_18} and low power Opamps.


\begin{table}[ht]
\centering
\caption{Comparison of Circuit Implemented using Resistive Threshold Logic with that of CMOS logic}
\begin{scriptsize}\begin{tabular}{p{2.5cm}p{1cm}p{1cm}p{1cm}p{1cm}}\hline\hline
Logic families & \multicolumn{2}{c}{16 bit full adder}& \multicolumn{2}{c}{16x1 MUX}\\
&Power & Area& Power & Area\\\hline
CMOS logic&2.5nW&4.557$\mu$m$^{2}$&0.189nW&1.070$\mu$m$^{2}$\\
Resistive logic (Opamp threshold)&3.277mW&8.081$\mu$m$^{2}$&0.447mW&0.825$\mu$m$^{2}$\\
\hline
\multicolumn{5}{p{8cm}}{Note:The power dissipation for  Opamps in the 16 bit full adder  is 2.47mW.  }

\end{tabular}
\end{scriptsize}\label{table71}
\end{table}

\section{Conclusion}
The concept of resistive  threshold logic was presented in an application to implement conventional digital logic gates.
 The presented resistive threshold logic family due to its ability to support large number of inputs can significantly help reduce the design complexity. Although, the presented resistive threshold outperforms the conventional CMOS logic  implementations in large input gates in terms of performance parameters such as area, delay and power, for small input gates further developments on low power and high speed Opamp designs are required.  The CMOS - Resistance Threshold Logic co-design  can optimise the circuit design of conventional CMOS based large variable boolean logic problems. A disadvantage of the proposed threshold logic using the  memristor technology in \cite{ref_14} as compared with CMOS logic is the higher power dissipation. However, with the advancements of newer low power memresitive devices such as \cite{ref_18}, the problem of lowering power dissipation  to the levels of CMOS, can be a realistic task. The  proposed logic can be extended to  technologies such as carbon  nanotubes and organic circuits. In addition, the ability of the proposed logic to develop large number of input gates can be seen as an early step in achieving the goal of mimicking brain like large variable boolean logic applications in VLSI.


%

%

\section*{Acknowledgment}
The authors would like to thank the anonymous reviewers for their time and thoughtful review comments, which has resulted in the improvement of overall quality of the brief.

\end{document}